\documentclass[preprint,authoryear,12pt]{elsarticle}

\usepackage{graphicx}
\usepackage{amssymb}

\usepackage{multirow}

\journal{Advances in Space Research}

\begin{document}

\begin{frontmatter}



\title{Short Term Topological Changes of Coronal Holes Associated with Prominence Eruptions and Subsequent CMEs
}


\author{H. Guti\'errez\corref{cor}\fnref{footnote1}}
\address{Space Research Center, University of Costa Rica}
\cortext[cor]{Space Research Center, University of Costa Rica, 2060 San Jos\'e, Costa Rica}
\fntext[footnote1]{Phone: +506 2511 6566, Fax: +506 2511 5147}
\ead{heidy.gutierrez@ucr.ac.cr}


\author{L. Taliashvili\corref{cor}\fnref{footnote2}}
\address{Space Research Center, University of Costa Rica}
\fntext[footnote2]{Phone: +506 2511 5099, Fax: +506 2511 5147}

\author{Z. Mouradian\corref{corr}\fnref{footnote3}}
\address{Observatoire de Paris--Meudon, LESIA}
\cortext[corr]{Observatoire de Paris--Meudon, LESIA}
\fntext[footnote3]{Phone: +33 1 45 07 78 00, Fax: +33 1 45 07 79 59}

\begin{abstract}
We study the short--term topological changes of equatorial and polar coronal hole (CH) boundaries, such as a variation of their area and disintegration, associated to reconnection with nearby (within 15$^\circ$ distance) quiescent prominence magnetic fields leading to eruptions and subsequent Coronal Mass Ejections (CMEs). The examples presented here correspond to the recent solar minimum years 2008 and 2009. We consider a temporal window of one day between the CH topological changes and the start and end times of prominence eruptions and onset of CMEs. To establish this association we took into account observational conditions related to the instability of prominence/filaments, the occurrence of a CME, as well as the subsequent evolution after the CME. We found an association between short--term local topological changes in CH boundaries and the formation/disappearance of bright points near them, as well as, between short--term topological changes within the whole CH and eruptions of nearby quiescent prominences followed by the appearance of one or more CMEs.
\end{abstract}

\begin{keyword}
coronal hole; coronal mass ejection; filament; magnetic field
\end{keyword}

\end{frontmatter}

\parindent=0.5 cm


\section{Introduction}

Coronal holes (CHs) are areas which are seen dark in X--rays and extreme ultraviolet \citep{waldmeier1,bohlin1,wang1}. Their plasma temperature and density is much lower than that of the ambient corona; so, they are low emission zones \citep{harvey2,raju1}. One of the important characteristics of CHs is their magnetic field, which is characterized by being mainly unipolar \citep{wang1,harvey2,raju1} with open magnetic field lines \citep{bohlin1,wang1,harvey2}; this means that its direction is primarily radial and the tangential component can be considered zero for distances greater than 2.5Rs \citep{wang3}, with Rs being the solar radius.  Closed magnetic field lines can keep the plasma within coronal loops; however, since the field is open in CHs, the plasma can escape into the interplanetary medium making up the solar wind.   It has been shown that the emergence of a small bipolar region from below the solar surface and its interaction with the preexisting open field in the coronal hole is a prime candidate to trigger reconnection and the consequent launching of jets along field lines in the corona \citep{moreno}.

Magnetic reconnection, which may occur between open and closed magnetic field lines or between open lines \citep{kahler2} is of great importance in the study of CHs, especially near their boundaries.  The process of continuous reconnection at CH boundaries is known as ``interchange reconnection'' \citep{wang4,wang1,fisk2,raju1,edmondson}.  \citet{wang2} have proposed two kinds of interchange reconnection, the first occurs when two open field regions of the same polarity are separated by photospheric flux of the opposite polarity and the reconnection takes place between the open flux domains and the underlying pair of loop systems. The second involves stepwise displacements within a region of single magnetic polarity and occurs when an open field line exchanges footpoints with a closed field line rooted next to it and the reconnection happens at the apex of the closed loop. Hence, coronal magnetic reconnection near CH boundaries can be responsible for their shape, magnetic topology and evolution.  Moreover, the appearance/disappearance of bright points (BPs) associated with short term evolution of CH boundaries \citep{nolte1,davis1,kahler1,madjarska1}, as well as CH topological changes associated with CMEs \citep{gonzalez1} or with filament eruptions and subsequent CMEs \citep{bravo1,gopalswamy2,jiang1,taliashvili2} reported in previous studies can be explained by the associated magnetic field configuration.

Regarding the origin of CMEs, statistical studies show that 70\% of prominence eruptions are associated with CMEs \citep{munro,pojoga}, moreover CHs may also be considered as signatures of CMEs \citep{bravo2,thompson}. In addition, filament eruptions are associated with the formation of small adjacent CHs and/or topological fluctuations of CHs \citep{harvey1,taliashvili1,taliashvili2}; and/or increased CH area \citep{bravo2,taliashvili1,taliashvili2}; and/or decrease and/or its complete disappearance \citep{taliashvili1,taliashvili2}. Using H$\alpha$ movies and spectroheliograms taken at Observatoire de Paris-Meudon, \citet{taliashvili2} reviewed 42 quiescent solar filaments/ prominences eruptions during two minimum periods of solar activity (1985--1986 and 1994). These authors found that the majority of prominence/filament eruptions ($\sim$91\%) are associated with the presence of adjacent CHs and subsequent CMEs. Magnetic reconnection is probably the mechanism responsible for the interaction between CHs and prominences located close to their boundaries, which may result in the onset of CMEs. However, the characteristics of this process is not yet clear. Recent studies show that nearly 15$^\circ$ distance between the prominences and CH boundaries could be considered as a possible critical distance for their interaction \citep{taliashvili2}.

The Sun is the primary driver of space weather; the most severe solar activity occurs around the cycle maximum, but its influence on the magnetosphere and ionosphere continues through the solar minimum \citep{cole}. Major events occur due to CMEs and high-speed solar wind streams associated with CHs \citep{feldman,echer,cole,schwenn}, as the solar wind speed originating in CHs is of about 750--800 km/s \citep{wang1,raju2}. It is important to note that both the CHs as well as their offspring, the high-speed solar wind streams, are representative of an inactive or quiet Sun \citep{feldman2,bame,schwenn}.  

Recent studies of the slow decline of solar Cycle 23 and slow rise of Cycle 24 show that the low solar activity lasted from about 2006 to the end of 2009, with 2008 and 2009 being particularly quiet years \citep{toma}, which is reflected in several solar irradiance observations, such as solar UV and EUV irradiance, and radio flux at 10.7 cm wavelengths \citep{tsurutani,solomon,hathaway}. Moreover, in 2007 and 2008, the effect of multiple, large, and long-lived CHs resulted in regular and recurrent solar wind streams and fast wind above 550 km/s, whereas in 2009, the disappearance of the low-latitude CHs shifted the sources of the solar wind to higher latitudes, mostly to the edges of the polar CHs with a drop of about 20\% in the mean speed and the almost total disappearance of fast solar wind. Generally, the continuous presence of multiple, low--latitude CHs during 2007 and 2008 and the very low magnetic flux emergence, made the minimum between Cycles 23 and 24 different from the two previous minima \citep{toma}.

In this work we study the possible association between CH topological changes and nearby prominence eruptions followed by CMEs as well as by sequences of CMEs. We analyze three events involving the evolution of the boundaries of CHs and of their surrounding regions (within $15^\circ$ distance), taking a temporal window of 12 days. We identify their topological changes from one day before until one day after a CME occurrence. The events took place during the recent solar minimum years, 2008 and 2009. The short--term topological changes of studied CHs are accompanied by the disappearance of nearby quiescent filaments, as well as subsequent CMEs.

\section{Data sources and working method}

The method used in this work is similar to the one reported by \citet{taliashvili1,taliashvili2}. We consider a period of 12 days of continuous observations of a CH evolution (far from big flares) including the pre- and post-CME stages; thus, we study CMEs not associated to flares. The selected quiescent filaments/prominences are located close to CH boundaries or its surrounding regions (within 15$^\circ$ distance). Our study comprises from August 24 to September 4 of 2008, December 3--14, 2008 and May 20--31, 2009. During these time periods we study one equatorial and 4 polar CHs, 3 filaments and 7 associated CMEs. This method allows us to identify every CH topological change before and after a non-flare CME.

The identification and daily evolution of each coronal hole is based on X--ray images supplied by Hinode (XRT, http://darts.isas.jaxa.jp/solar/hinode/), EUV images from EIT/SOHO (http://sohowww.nascom.nasa.gov/), EUVI/\-STEREO (http://stereo.gsfc.nasa.gov/), SECCHI/STEREO movies (http:// secchi.nrl.navy.mil/index.php?p=movies) and  magnetograms from the Wilcox Solar Observatory (WSO, http://wso.stanford.edu/) and Michelson Doppler Imager (MDI) instrument on the SOHO (http://sohowww.nascom. nasa.gov/). 

The analysis of quiescent filaments located close to a CH boundary is done using observations from the Global High Resolution H--alpha Network (http://swrl.njit.edu/ghn\_web/), EUV images ($\lambda=$304$\rm{\mathring A}$) obtained by EIT/ SOHO and EUVI/STEREO, and SECCHI/EUVI--304$\rm{\mathring A}$ synoptic maps (http:// secchi.nrl.navy.mil/synomaps/).

We identify the associated CMEs considering the position angles of CHs, prominences and CMEs; the starting times of prominence eruptions and CMEs and, CH topological changes within 1 day before and after a CME. CMEs are identified using CME catalogs, images and movies from STEREO/ COR1 and SOHO/LASCO.

\section{Results}

\subsection{Event 1. August, 24--September, 04 2008. CH1+F1}

During this period we observe a polar coronal hole (CH1), located between N90$^\circ$--N60$^\circ$,  with negative polarity (based on WSO magnetograms) and a maximum latitudinal width of about 30$^\circ$. The filament (F1) at $\sim$N55$^\circ$ (at least 30$^\circ$ long) extends at a distance of $\sim$10$^\circ$, almost parallel to the southward boundary of CH1. 

\begin{table}
\caption{CMEs associated with the events under study. CMEs reported for Event 2 correspond to the same eruption.}{\footnotesize
\begin{tabular}{p{1.2cm}p{1.2cm}p{1.5cm}p{1.2cm}p{1.2cm}p{1cm}p{3cm}p{1.8cm}}
\hline
{\bf Event} & {\bf CME} & {\bf Date mm/dd} & {\bf Onset Time (UT)} & {\bf Position Angle (PA)} & {\bf Width (W)} & {\bf Source} & {\bf Figure} \\
\hline
\multirow{3}{4cm}{Event 1} & CME1.1 & 2008/08/29 & 14:06 & $\sim$300$^\circ$ & $\sim$26$^\circ$ & STEREO-B/COR1 & Figure 4.a \\
 & CME1.2 & 2008/08/29 & 21:30 & $\sim$300$^\circ$ & $\sim$20$^\circ$ & STEREO-A/COR1 & Figure 4.b \\
 & CME1.3 & 2008/08/30 & 16:06 & $\sim$300$^\circ$ & $\sim$22$^\circ$ & STEREO-B/COR1 & Figure 4.c \\
\hline
\multirow{3}{4cm}{Event 2} & CME2.B & 2008/12/12 & 04:35 & $\sim$310$^\circ$ & $\sim$40$^\circ$ & STEREO-B/COR1 & Figure 8.a\\
 & CME2.A & 2008/12/12 & 04:55 & $\sim$50$^\circ$ & $\sim$35$^\circ$ & STEREO-A/COR1 & Figure 8.c \\
 & CME2.L & 2008/12/12 & 08:54 & $\sim$297$^\circ$ & $\sim$115$^\circ$& LASCO/SOHO & Figure 8.b\\
\hline
\multirow{3}{4cm}{Event 3} & CME3.1 & 2009/05/29 & 03:45 & $\sim$245$^\circ$ & $\sim$30$^\circ$ & STEREO-B/COR1 & Figure 11.a \\
 & CME3.2 & 2009/05/29 & 09:30 & $\sim$258$^\circ$ & $\sim$74$^\circ$ & LASCO/SOHO & Figure 11.b \\
& CME3.3 & 2009/05/29 & 15:10 & $\sim$242$^\circ$ & $\sim$23$^\circ$ & STEREO-A/COR1 & Figure 11.c \\
\hline
\end{tabular}}
\label{table1}
\end{table}

F1 becomes unstable that starts disappearing on August 29 at 04:04 UT. This disappearance lasts $\sim$35 hours based on data from STEREO--A/EUVI--304$\rm{\mathring A}$ and STEREO--B/EUVI--304$\rm{\mathring A}$. Twelve hours before and $\sim$5h after the starting time of F1 disappearance, SOHO/LASCO reports two narrow CMEs, at Position Angle (PA) = 355$^\circ$, Width (W) = 9$^\circ$ and PA=357$^\circ$, W=9$^\circ$, respectively. Due to their small widths (W$<$10$^\circ$) we are not including this type of narrow CMEs in our study. Additionally, $\sim$10h and $\sim$36h after the starting time of F1 disappearance, STEREO--B/COR1 reports two CMEs, CME1.1 (at 14:06 UT) and CME1.3 (at 16:06 UT), and STEREO--A/COR1 reports CME1.2 (21:30 UT). Based on STEREO--A/B/COR1 observations we estimate the PA and W for CME1.1, CME1.3 and CME1.2, this information is shown in Table \ref{table1}. The spatial location and time of these three CMEs are well correlated with the long duration disappearance of F1. The disappearance of F1 is accompanied by changes in the nearby polar coronal hole. 

\begin{figure}
  \begin{center}
 \includegraphics*[width=13cm]{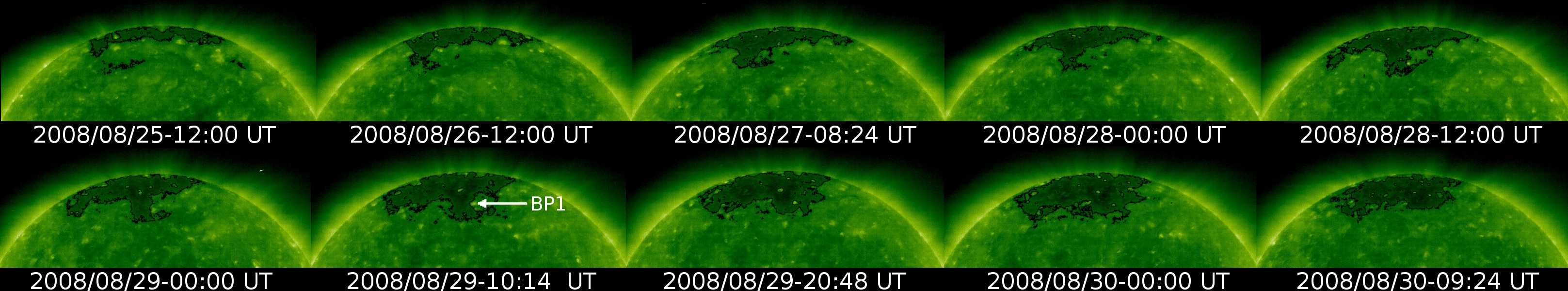}
\end{center}
\caption{EIT/EUV--195$\rm{\mathring A}$ image sequence of CH1 from August 25 to 30 showing BP1 near to its western boundary. We have enhanced the CH boundaries for clarity.}
\label{figure1}
\end{figure}

\begin{figure}
\begin{center}
 \includegraphics*[width=13cm,]{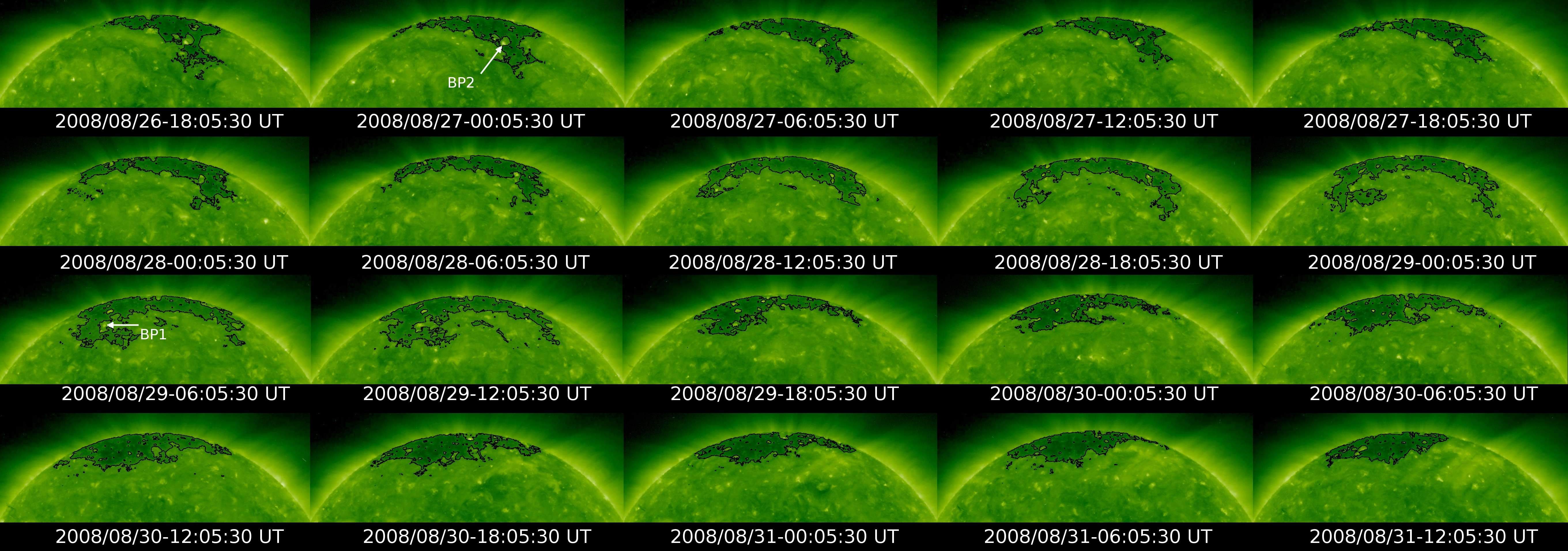}
\end{center}
\caption{STEREO--A/EUVI--195$\rm{\mathring A}$ image sequence from August 26 to 31 showing shape and size changes of CH1. The white arrow indicates BP1 and BP2 near to the eastern and western boundaries (of the western and eastern sections, respectively) of CH1. We have enhanced the CH boundaries for clarity.}
\label{figure2}
\end{figure}

\begin{figure}
\begin{center}
 \includegraphics*[width=13cm]{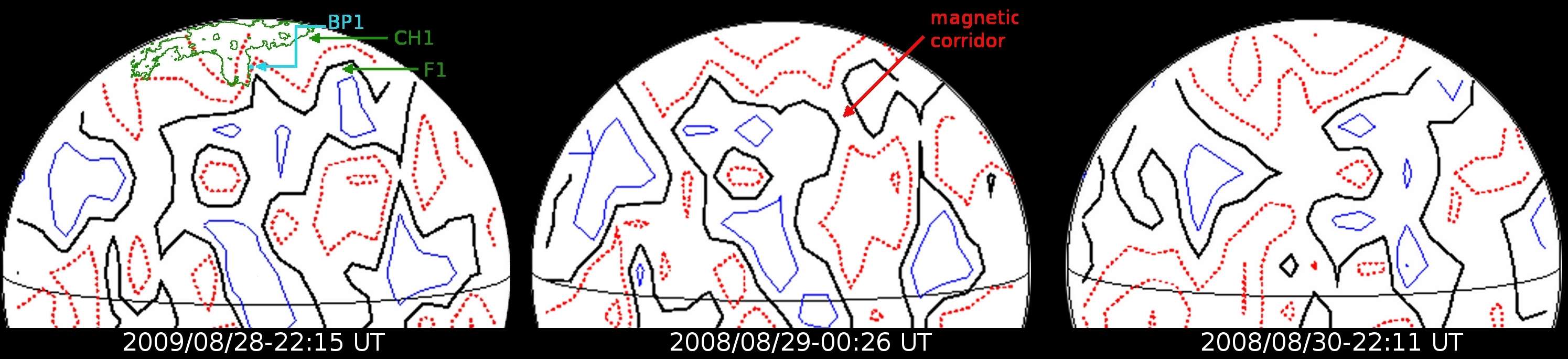}
\end{center}
\caption{WSO magnetograms showing the magnetic corridor and magnetic configuration associated with BP1, the eastern section of CH1 and F1. BP1, CH1 and F1 are indicated by arrows. WSO magnetic contour maps show the regions of positive and negative polarity as blue and red areas.}
\label{figure3}
\end{figure}

A small section of CH1 that is located close to central meridian in EIT (Figure \ref{figure1}; visible at the east by STEREO--A, Figure \ref{figure2}), starts to increase in latitudinal width one day before F1 disappearance (on August 28, 00:00 UT. Figure \ref{figure1} and August 28, 00:05 UT. Figure \ref{figure2}) and extends up to $\sim$N50$^\circ$ during the next day; moreover, its longitudinal width is $\sim$40$^\circ$ (August 29, 06:05 UT. Figure \ref{figure2}). These changes last for $\sim$8h (August 29, 12:05 UT. Figure \ref{figure2}). Subsequently, this section begins to decrease in latitudinal width and reaches again $\sim$N60$^\circ$ (August 30, 00:05 UT. Figure \ref{figure2}), whereas its longitudinal width continues to grow (up to $\sim$110$^\circ$) (August 30, 18:06 UT. Figure \ref{figure2}), which leads to the growth of the area of this section (as shown on August 30, 00:00 UT in Figure \ref{figure1} and August 30, 12:05 UT in Figure \ref{figure2}). The starting and ending times of this process approximately coincide with the onset of CME1.1 and CME1.3 respectively. The consecutive growth and decrease in latitudinal width and the growth in longitudinal width of this eastern section of CH1 coincides with the appearance and disappearance of bright point (BP1) near its western boundary (it appears on August 28, $\sim$07:00 UT and disappers on August 30, $\sim$01:30 UT, see for example Figure \ref{figure2} on August 29, 06:05 UT) and with the eruption of filament (F1) in progress.  BP1 is associated with a small magnetic dipole observed by SOHO/MDI that disappears almost simultaneously with BP1.  Similar observations, such as X-ray bright points associated with pairs of opposite-polarity photospheric magnetic fragments, have been reported by several authors (for example, \citet{priest}). BP1 lifetime agrees with the short--term magnetic reorganization, specially from  August 28 (22:15 UT) to 29 (00:29 UT), during which the negative polarity CH1 grows and subsequently, $\sim$3h before F1, a magnetic corridor with negative polarity forms between the polar and equatorial regions that extends from BP1 surrounding, near the F1 (Figure \ref{figure3}). Whereas, the western part of CH1 (that is only seen by STEREO--A, Figure \ref{figure2}) decreases from August 26, followed by the appearance of BP2 close to its eastern boundary. This BP2 disappears almost simultaneously with the start of the eruption of F1 and with the process that initiates the decrease of the western section of CH1. After the onset of CME1.2, this section disappears completely almost simultaneously. Due to the position of BP2 at the far side of the Sun, we can not study its magnetic evolution. Therefore, the total CH1 area decreases continuously from August 27 reaching its minimum value on August 28 ($\sim$24h before the onset of F1 disappearance), later it increases again until August 29 (approximately 10h after the onset of F1 eruption and close to the onset of CME1.1).  Subsequently, the total area of CH1 decreases slightly just before the onset of CME1.1 and CME1.2. The period after the onset of the CMEs on August 29 coincides with the growth in longitudinal width of CH1, which reaches its maximum width few hours after CME1.3 starting time. These variations of the area of CH1 are accompanied by morphological variations, from an irregular shape to an elliptical one.  The magnetic corridor disappears and the magnetic configuration recovers after the disappearance of BP1 and CMEs (Figure \ref{figure3}).

\begin{figure}
\begin{center}
\includegraphics*[width=13cm]{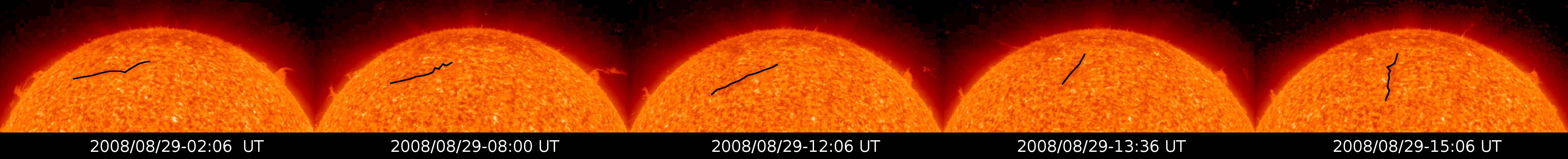}
\end{center}
\caption{STEREO--A/EUVI--304$\rm{\mathring A}$ images showing the temporal and spatial evolution of F1 eruption. The whole filament F1 is observed on the first image (August 29, 02:06 UT). The images that follow show the direction of the eruption of F1 that is hand--drawn mainly based on STEREO--A/EUVI--304$\rm{\mathring A}$ movies.}
\label{figure4}
\end{figure}

\begin{figure}
\begin{center}
 \includegraphics*[width=13cm]{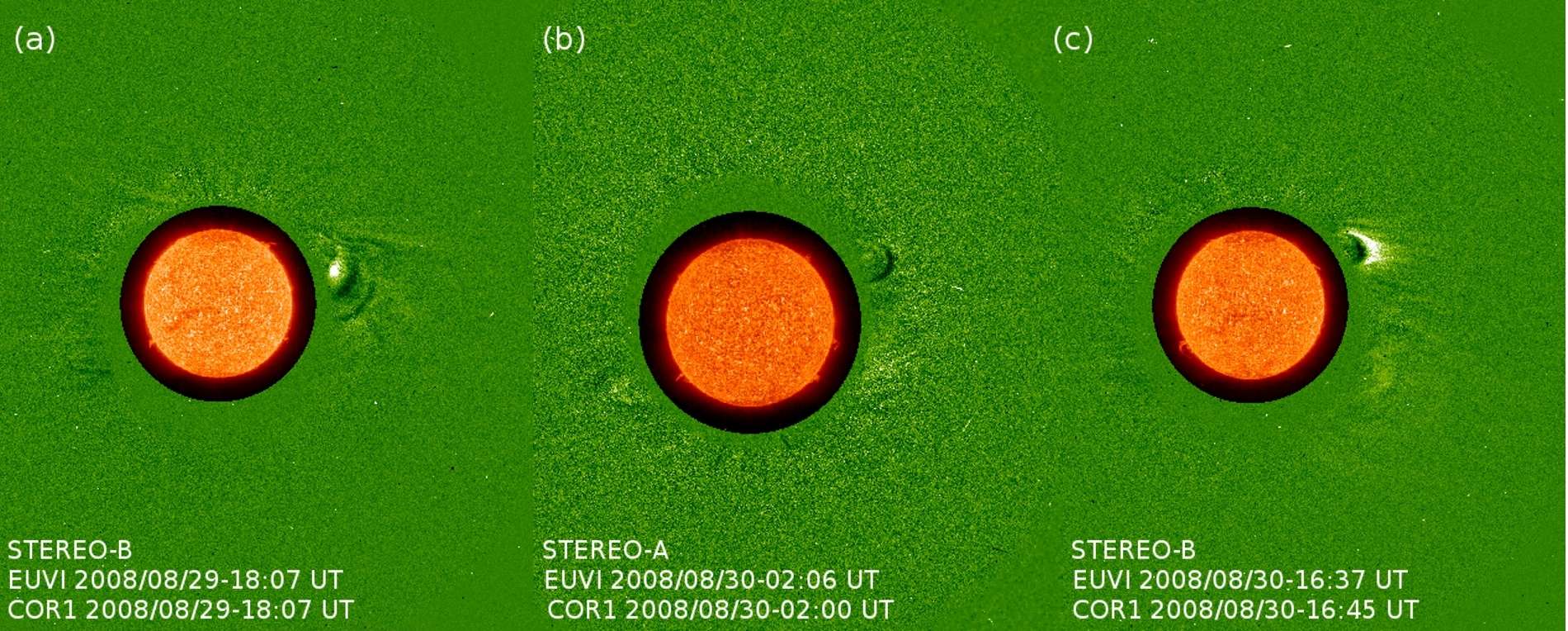}
\end{center}
\caption{Composition of STEREO images observed by COR1 and EUVI--304$\rm{\mathring A}$. {\bf(a)}, {\bf(b)} and {\bf(c)} show CME1.1, CME1.2 and CME1.3 respectively. CME1.1 and CME1.2 are deflected southward from their radial position.}
\label{figure5}
\end{figure}

F1 disappearance starts in the eastern hemisphere of the Sun as observed by STEREO--A and appears as a west limb eruption when viewed by STEREO--B. The eruption of F1 is disperse and its direction crosses the solar disk westward. This non--radial eruption propagates in two directions from STEREO--A perspective (Figure \ref{figure4}). The primary direction is southward then it moves away from the eastern part of CH1 and curves westward or to the western CH1 boundary. The motion of F1 coincides with the decrease of the western part of CH1 and the growth of the eastern part of CH1. The maximum area of eastern section and the minimum of the weastern section, are observed simultaneously at the ending time of F1 eruption. The primary direction of F1 is well-correlated with CME1.1 (STEREO--B) that is followed after 7.5h by CME1.2 (STEREO--A). The position angles of both CMEs are displaced $\sim$20$^\circ$ southward from F1 position angle. The propagation of the CME is not radial in either case, and in the second is also deflected southward. Regarding the last CME1.3 that starts $\sim$36h after the starting time of F1 eruption (STEREO--B), we consider that it is associated to the eruption of the remnant portion of F1 due to the well correlated position angles of CME1.3 and a small section of F1. However, we also consider that CME1.3 could be associated to another activity on the far side of the Sun, which is impossible to observe since the separation between STEREO--A/B (70$^\circ$) is insufficient. In general, three West limb CMEs observed by both COR1--B (CME1.1, Figure \ref{figure5}.a and CME1.3, Figure \ref{figure5}.c) and COR1--A (CME1.2, Figure \ref{figure5}.b) are in good correlation with the moving F1 prominence material.

\subsection{Event 2. December 3--14, 2008. CH2+CH3+F2}

During this period we analyze a system formed by one polar (CH2) and one equatorial coronal hole (CH3) separated by $\sim$10$^\circ$. CH2 is located between N90$^\circ$ -- N50$^\circ$, with a maximum of $\sim$40$^\circ$ latitudinal width; CH3 is located at  $\sim$N30$^\circ$, with a maximum of $\sim$18$^\circ$ longitudinal and $\sim$25$^\circ$ latitudinal widths, both on negative polarity field. Moreover, the filament (F2), at $\sim$N55$^\circ$ (at least $\sim$30$^\circ$ length) extends from $\sim$N50$^\circ$ to $\sim$N35$^\circ$, between both coronal holes, with a maximum separation of $\sim$10$^\circ$ eastward of the CHs.

F2 starts erupting on December 12 at 00:06 UT, this eruption lasts $\sim$8 hours based on observations of STEREO--A/EUVI--304$\rm{\mathring A}$ and STEREO--B/EUVI--304$\rm{\mathring A}$. About 4.5h after the starting time of F2 disappearance, STEREO--B/COR1 reports CME2.B (at 04:35 UT) and after $\sim$20 min, STEREO --A/COR1 reports the same CME, CME2.A (at 04:55 UT). Based on COR1/ STEREO--B/A image analyses we estimate the PA and W for both CMEs (see Table \ref{table1}).  Additionally, about 1h after the ending time of F2 disappearance, a CME starts, reported by SOHO/LASCO (CME2.L, at 08:54 UT, PA=297$^\circ$, Table \ref{table1}); based on LASCO images, it expands and reaches W$\sim$115$^\circ$ at 10:54h. We consider that these three CMEs correspond to the same CME (CME2) observed by LASCO and COR1 aboard STEREO--A/B. The spatial location and time appearance of these LASCO/STEREO CMEs are well correlated with the long duration F2 disappearance.

\begin{figure}
\begin{center}
 \includegraphics*[width=13cm]{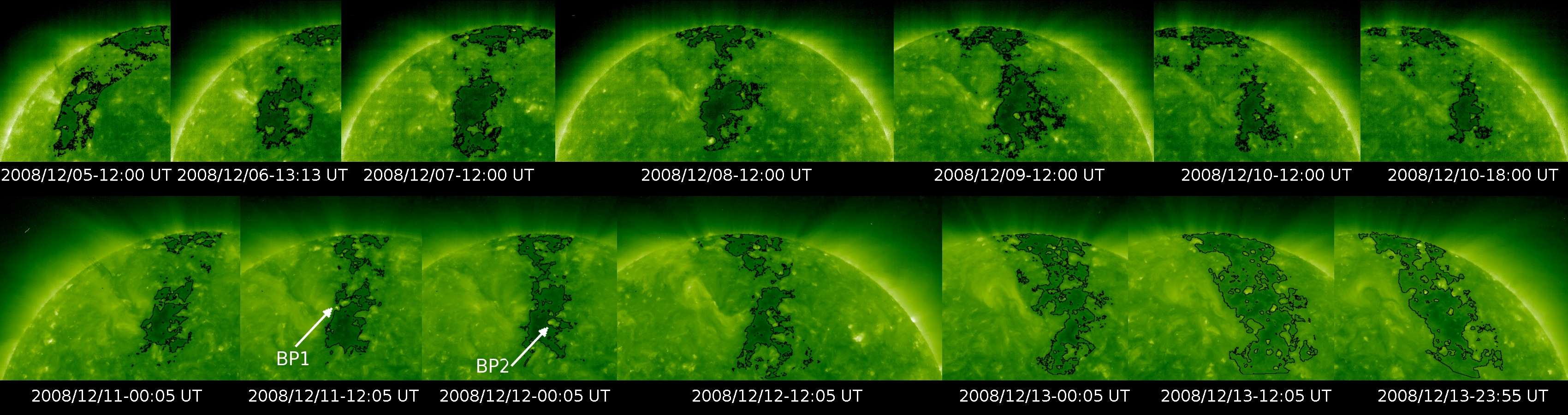}
\end{center}
\caption{Temporal and spatial evolution of CH2 and CH3 during a 9--days period, observed by EIT/EUV--195$\rm{\mathring A}$ (top) and STEREO--A/EUVI--195$\rm{\mathring A}$ (bottom) images. The white arrows indicate bright points on CH3 boundary. We have enhanced the CH boundaries for clarity.}
\label{figure6}
\end{figure}

On December 05 we observe two coronal holes, CH2 and CH3 separated by $\sim$25$^\circ$ (Figure \ref{figure6}) that start merging, during the next four days their separation reduces up to $\sim$5$^\circ$ (December 9, 12:00 UT), later they separate again. About 23h before F2 disappearance, the separation between both CHs is maximum (December 11, 00:05 UT. Figure \ref{figure6}). Simultaneously, close to the eastern boundary of CH3 a BP, BP1, appears (December 11, 12:05 UT. Figure \ref{figure6}) and stays there for less than a day.  Moreover, just a few hours before the starting time of F2 disappearance, another BP, BP2, (December 12, 00:05 UT. Figure \ref{figure6}) is formed near by, which starts weakening after the end of the filament eruption (and after CME2) and is not observed within the next $\sim$23h; at the same time, small sections of CH3 around BP1 appear fragmented. In addition, at the time F2 disappearance begins, both CHs start to merge and form a single coronal hole (December 12, 00:05 UT. Figure \ref{figure6}). This reconnection occurs close to the location of BP1 that disappears during this process. Later, $\sim$12 hours after the onset of F2 disappearance (and after CME2), the connection between the CHs starts to disappear (December 12, 12:05 UT. Figure \ref{figure6}), but within the next 12h they form a single coronal hole for a second time. This CH starts to expand (December 13, 00:05 UT. Figure \ref{figure6}), reaches its maximum area on December 13 (12:05 UT Figure \ref{figure6}) and remains stable during the two following days. BP2 formation and disappearance, agrees in time with the CH disconnection and reconnection during December 12.  Each of bright points (BP1 and BP2) is associated with a small magnetic dipole observed by SOHO/MDI that disappears simultaneously with BP1 and BP2 respectively.

\begin{figure}
\begin{center}
 \includegraphics*[width=13cm]{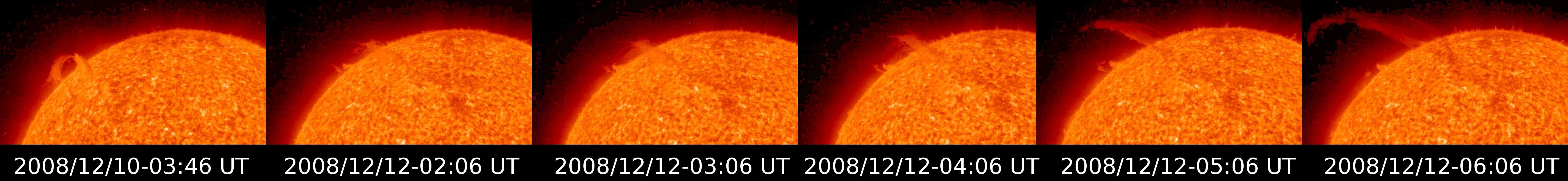}
\end{center}
\caption{F2 Eruption observed by STEREO--A/EUVI--304$\rm{\mathring A}$. The entire prominence is observed in the first image (December 10, 03:46 UT). The following images show F2 eruption. }
\label{figure7}
\end{figure}

\begin{figure}
\begin{center}
 \includegraphics*[width=13cm]{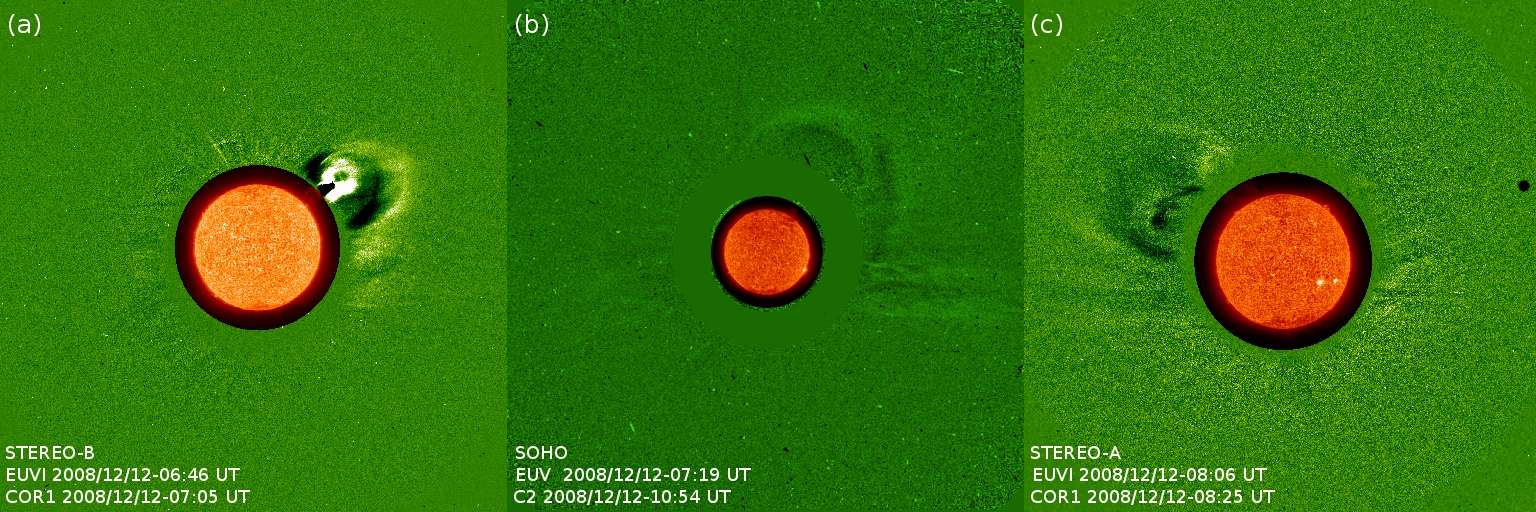}
\end{center}
\caption{CME2 observed by the composition of STEREO--B/A images ({\bf a}, {\bf c}, respectively) from COR1 and EUVI--304$\rm{\mathring A}$ and SOHO images {\bf(b)} from LASCO--C2 and EIT/EUV--304$\rm{\mathring A}$. {\bf (c)} clearly displays the CME deflection southward from its radial direction.}
\label{figure8}
\end{figure}

Figure \ref{figure7} shows the evolution of F2 eruption in STEREO--A/EUVI--304$\rm{\mathring A}$ images. The trajectory of this eruption is not radial and moves away from CH2+CH3. Similarly, the direction of CME2.A observed by COR1--A is not radial and is deflected southward (Figure \ref{figure8}.c). The position angle of CME2.A is displaced $\sim$20$^\circ$ southward from F2 position angle. However, STEREO--B/EUVI--304$\rm{\mathring A}$ does not detect the deflection of F2. The CME deflection is lower in COR1 B and LASCO (Figure \ref{figure8}.a and \ref{figure8}.b). \citet{panasenco1} have analyzed the deflection of this CME.

\subsection{Event 3.  May 20--31, 2009. CH4+F3}

During this period we study one polar,  positive polarity CH (CH4), reaching S45$^\circ$ and a polar filament F3 at $\sim$S50$^\circ$ (at least 35$^\circ$ long) that extends almost parallel to the equator, with a maximum of $\sim$15$^\circ$ separation from the southern boundary of CH4. 

F3 starts disapeparing on May 29 at 02:06 UT, this process lasts $\sim$14h based on 304$\rm{\mathring A}$/STEREO--A and 304$\rm{\mathring A}$/STEREO--B observations. About 30min before and $\sim$7.5h after F3 disappearance starting time, two LASCO/CMEs are observed, narrow CME at 01:31 UT (PA=211$^\circ$ and W=8$^\circ$) and CME3.2 at 09:30 UT (see Table \ref{table1}). As for the first event, we do not consider the narrow CME1 in this study. Additionally, $\sim$1.75h and $\sim$13h after the starting time of F3 disappearance, COR1/STEREO--B and COR1/STEREO--A report CME3.1 (at 03:45 UT) and CME3.3 (at 15:10 UT). Based on the corresponding COR1/STEREO--A/B images, we estimate their PA and W in Table \ref{table1}. Additionally, we include CME3.2 (observed by LASCO) and CME3.3 (observed by STEREO--A) a continuation of CME3.1 (observed by STEREO--B). The spatial location and time of these three CMEs is well correlated with the long duration F3 disappearance that is accompanied by variations in the nearby polar coronal hole.

\begin{figure}
\begin{center}
 \includegraphics*[width=13cm]{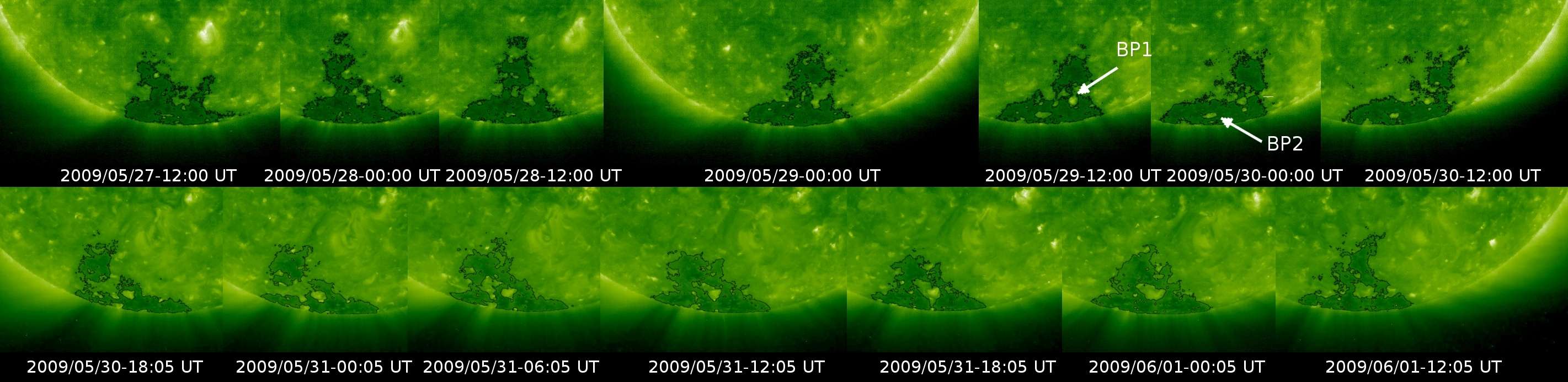}
\end{center}
\caption{Temporal and spatial evolution of CH4 observed by EIT/EUV--195$\rm{\mathring A}$ (top) and STEREO--A/EUVI--195$\rm{\mathring A}$ (bottom) through six days. The white arrows indicate bright points in CH4. We have enhanced the CH boundaries for clarity.}
\label{figure9}
\end{figure}

On May 27 a section of CH4, located at $\sim$S47$^\circ$ starts to grow toward the equator, reaching $\sim$S37$^\circ$ (May 28, 12:00 UT. Figure \ref{figure9}) and from the next day on it decreases up to $\sim$S45$^\circ$ (May 29, 00:00 UT. Figure \ref{figure9}) just 2h before the starting time of F3 disappearance. The CH stays at this latitude during the next few hours. Two associated bright points emerge, first BP1 on May 28 on the western border of CH4 that is best seen after CME3.2 (May 29, 12:00 UT. Figure \ref{figure9}); the second BP2 appears at approximately the same time as the onset of CME3.1 (May 30, 00:00 UT. Figure \ref{figure9}). About 12h after the onset of F3 disappearance ($\sim$1h before the onset of CME3.3), BP1 intensifies and simultaneously CH4 fragments in two main sections around this BP1; the maximum separation is observed in the early hours on May 31 (00:05 UT. Figure \ref{figure9}). BP2 is close to  the southern boundary of the southern section that observed separately during the next $\sim$6h  (May 31, 00:05 UT. Figure \ref{figure9}) and then reconnects again (May 31, 06:05 UT. Figure \ref{figure9}). BP1 persists till June 02, while BP2 lasts until May 31. On the other hand, the growth of BP2 throughout May 29 is associated with a decrease of the entire CH4 area; this process continues accompanied by the fragmentation of the CH and lasts until May 31 (00:05 UT. Figure \ref{figure9}). After a few hours these fragments join again.  Due to the polar positions of the bright points, unfortunately we could not observe the magnetic configuration associated with them in magnetograms.

\begin{figure}
\begin{center}
\includegraphics*[width=13cm]{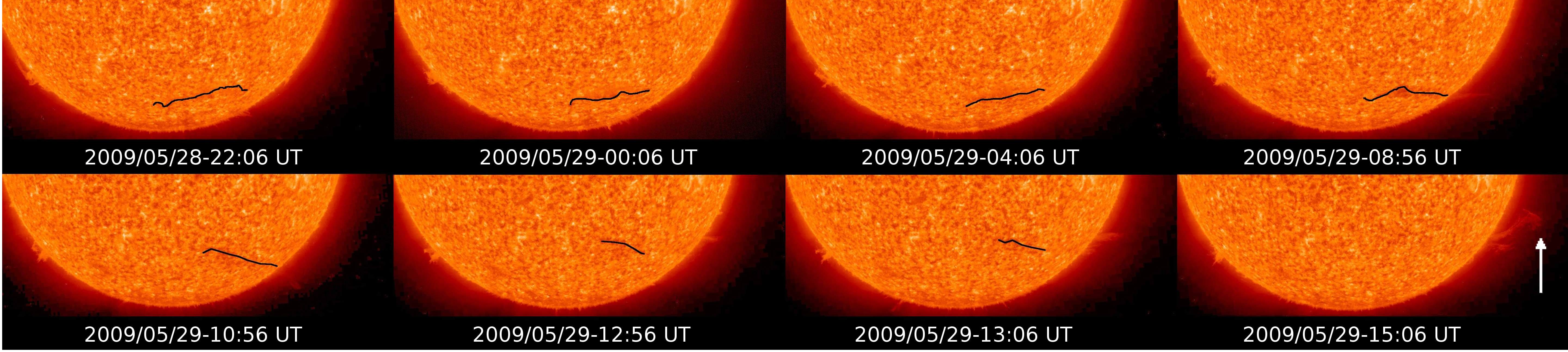}
\end{center}
\caption{Temporal and spatial evolution of F3 eruption observed by STEREO--A/EUVI--304$\rm{\mathring A}$. The entire filament F3 is observed on the first image (May 28, 22:06 UT). Following images show the direction of F3 eruption that is hand--drawn mainly based on SECCHI STEREO--A/EUVI--304$\rm{\mathring A}$ movies.}
\label{figure10}
\end{figure}

\begin{figure}
\begin{center}
 \includegraphics*[width=13cm]{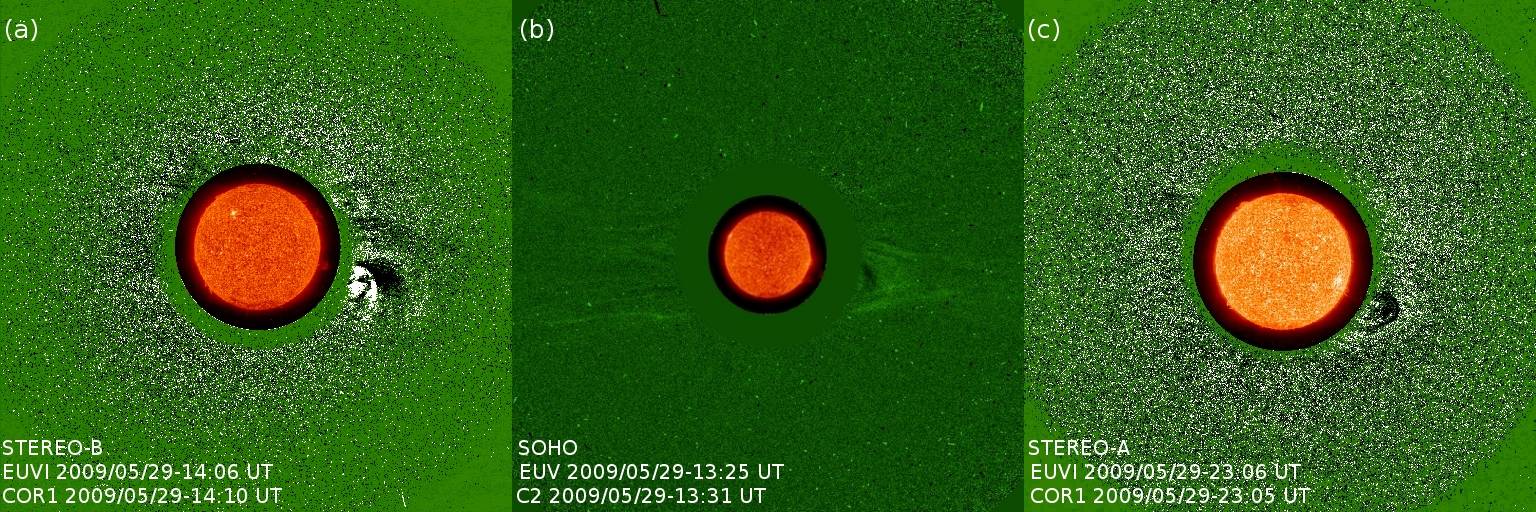}
\end{center}
\caption{Left and right composed STEREO images from COR1 and EUVI--304$\rm{\mathring A}$ and center composed SOHO images from by C2 and EIT/EUV--304$\rm{\mathring A}$. {\bf(a)} CME3.1 {\bf(b)} CME3.2 {\bf(c)} CME3.3. CME3.1 and CME3.3 deflect to northward from their radial positions.}
\label{figure11}
\end{figure}

F3 disappearance starts on the west hemisphere of the Sun observed by STEREO--A, and as a west limb prominence eruption viewed by STEREO--B. Figure \ref{figure10} shows the evolution of F3 disappearance/eruption observed by STEREO--A. F3 eruption crosses the solar disk to the NW, and after $\sim$7h is observed as a west limb prominence eruption, always moving toward the equator. The direction of the moving F3 eruption coincides with the STEREO/LASCO consecutive three west limb CMEs; first CME3.1 is observed by COR1--B, after $\sim$6h, the second LASCO/CME3.2 starts and finally, after $\sim$6h, the COR1--A/CME3.3. The eruption of F3 (as seen from STEREO--A) is deflected to the North (Figure \ref{figure10}). CME3.1 and CME3.3 do not follow a radial direction, they deflect northward (Figure \ref{figure11}.a and \ref{figure11}.c). The difference between the position angles of CME3.1 and the prominence is $\sim$15$^\circ$, whereas the one between CME3.3 and the prominence is $\sim$12$^\circ$. CME3.2 is not deflected and the difference between the position angles of CME3.2 and F3 is $\sim$23$^\circ$.

\section{Discussion and Conclusions}

We study the detailed evolution of three different systems composed by coronal holes and nearby (within 15$^\circ$ distance from their boundaries) quiescent filaments: CH1+F1, CH2+CH3+F2 and CH4+F3, that include polar CHs as well as equatorial ones. For each system we analyze the short--term CH topological changes during a period of 12 days associated with the disappearance of nearby filaments, especially before and/or after the disappearance. Moreover, we study the associated CMEs observed at similar position angles (in total 7 CMEs, excluding those with less than 10$^\circ$ width) as the studied prominences and nearby coronal holes.  

 For all the studied events we observed the ejection of CMEs after the onset of the filament eruptions, within 36 hours of the starting time of prominence eruptions. We observe that during a prominence eruption at the northern hemisphere, a section of the CH (event 1) or the entire CH (event 2), located in opposite direction to this eruption, grows in extension; whereas the other section during the event 1, which is located towards the moving erupting prominence, decreases. Additionally, for event 2, the increase of the CH areas is more pronounced, probably due to the merging of both CHs. For both events, we observe associated bright points (two BPs for each events) before the starting time of prominence eruptions and their disappearance after the eruption. In addition, at least one of the BPs is associated to the visible reduction and fragmentation of the CH. In the case of event 1, the BP (located at the eastern section of CH1) lifetime coincides with the growth of this eastern section and with the period of time between the onset of CME1.1 and CME1.3, whereas the disappearance of the BP (located at the western section of CH1) coincides with the decrease of the western section and with CME1.2 starting time; after the onset of CME1.2, this western section disappears completely. For event 2, the appearance of one of the BPs coincides with the maximum separation between both CHs (CH2 and CH3); then, it disappears after the CME2 onset, whereas the lifetime of the other BP coincides with the time period between the disconnection and reconnection of the CHs (CH2+CH3) and it disappears before the CME2 onset.  Regarding event 3, located at the South Pole, we observe the appearance of a BP (BP1) before the starting time of prominence eruption followed by the appearance of other BP (BP2) and two CMEs (CME3.1, CME3.2), then BP1 intensifies, simultaneously CH4 fragments in two main sections around this BP; this is followed by CME3.3. After these 3 CMEs the BPs disappear. Each of the bright points observed near the CH (event 1 and event 2) boundaries is associated  with a small magnetic dipole that appears and disappears simultaneously with BP.  

The most evident topological changes of the CHs are seen after the onset of the CMEs. For event 1 (North Pole CH), between the starting times of CME1.1 and CME1.3, the eastern section of CH1 (which lays in the same direction as CME1.1 and CME1.3) decreases its latitudinal width and increases its longitudinal width, whereas the small part of its western section disappears before the onset of the west limb CME1.2. For event 2, CH2 (North Pole CH) and CH3 (equatorial CH) appear separated after CME2 onset. Regarding event 3 (South Pole CH), CH4 disintegrates after CME3.2 onset, and after CME3.3 onset the separation between these segments reaches its maximum. 

We have found that short--term topological changes in the entire CHs are associated with nearby quiescent prominence eruptions, particularly when a separation between the quiescent filaments and the CH boundary is less than or equal to 15$^\circ$, followed by the ejection of one or more CMEs \citep{taliashvili2}.  Similar results regarding the separation distance (of $\sim$10$^\circ$) between the filament and CH boundaries, associated with prominence eruption and subsequent CME were reported recently by \citet{panasenco1}.

 In addition, our observations indicate that BP appearance or disappearance is related with the starting processes of nearby filament eruptions, which can be caused by disturbances of the CH environment. These disturbances are most likely the result of magnetic reconnection between magnetic field lines associated with BPs and the surrounding of CH boundaries followed by the reorganization the magnetic field, which results in different BP lifetimes and CH boundary or area changes and reaches the foot points of nearby filaments, contributing in this way to their destabilization, eruption, and subsequent CME ejection. There are several proposed models and observational evidences related to magnetic reconnection at (or close to) the CH boundaries  and related with visible appearance/disappearance of BPs, filament eruptions, short-term topological changes in the entire CHs and subsequent formation of CMEs \citep{kahler1,kahler2,bravo1,bravo2,madjarska1,wang2,fisk2}. Specially, magnetic bipoles emerging within a CH or small loops formed in the CH during interchange reconnection could increase along a CH lifetime and ultimately lead to the fragmentation and diffusion of the CH \citep{krista1}. Interchange reconnection at CH boundaries is supported by several observational results \citep{krista1}, e.g., the prevalence of very small loops inside CHs and larger loops outside CHs  \citep{wiegelmann} and the boundary displacements observed due to the emergence and disappearance of bright small--scale loops in the form of BPs \citep{madjarska2,subramanian}. 

In addition, we observe that the direction of the erupted filaments near CHs and associated CMEs is almost non--radial. Both structures (regarding the three studied events) are moving toward the equator. In addition, the difference between the central position angles of both ranges between 15$^\circ$ and 20$^\circ$. These observations are consistent with the results reported by \citet{gopalswamy1,cremades1,gopalswamy3,panasenco1};  they found that the CMEs generally move away from the open magnetic field regions and their the deflection is probably due to the fact that at lower coronal heights they are guided by the open field along which the fast solar wind flows.

 In this study we are not considering long--term topological variation of coronal holes in regard to different stages of the solar cycle and the associated involvement of prominence eruptions and subsequent CMEs, which we plan to pursue in an upcoming project.

\vspace{0.5cm}

{\footnotesize{\bf Acknowledgments.}We are grateful to the Hinode, STEREO, SOHO and Global High Resolution H--alpha Network for open access to their data sets. Hinode is a Japanese mission developed and launched by ISAS/JAXA, with NAOJ as domestic partner and NASA and STFC (UK) as international partner. It is operated by these agencies in co--operation with ESA and the NSC (Norway). LASCO and EIT are part of SOHO, SOHO is a project of international cooperation between ESA and NASA. The LASCO CME catalog is generated and maintained at the CDAW Data Center by NASA and The Catholic University of America in cooperation with the Naval Research Laboratory. The STEREO mission is supported by NASA, PPARC (UK), DRL (Germany), CNES (France), and USAF. The SECCHI data used here were produced by an international consortium of the Naval Research Laboratory (USA), Lockheed Martin Solar and Astrophysics Lab (USA), NASA Goddard Space Flight Center (USA), Rutherford Appleton Laboratory (UK), University of Birmingham (UK), Max--Planck--Institut for Solar System Research (Germany), Centre Spatiale de Li\`ege (Belgium), Institut d'Optique Theorique et Appliqu\'e (France), Institut d'Astrophysique Spatiale (France). The ``COR1 Preliminary Events List'' was generated by O. C. St. Cyr prior to September 2007, and is being maintained now by Hong Xie.  Wilcox Solar Observatory is currently supported by NASA and data used in this study was obtained via the web site, courtesy of J.T. Hoeksema. We are grateful to M. S\'anchez and T. Roinishvili to improve English. This study was performed as a partial requirement for the PhD Degree of Sciences at the University of Costa Rica.  Special thanks are owed to anonymous referees for constructive comments that helped to improve the quality of the paper.}


\section*{Bibliography}

\begin{thebibliography}{}

\bibitem[Bame et al.(1977)]{bame}
Bame, S.J., Asbridge, J.R., Feldman, W.C., Gosling, J.T.
Evidence for a Structure--Free State at High Solar Wind Speeds.
Journal of Geophysical Research 82, 1487--1492, 1977.

\bibitem[Bohlin(1977)]{bohlin1}
Bohlin, J.D.
Extreme--Ultraviolet Observations of Coronal Holes. I: Locations, Sizes and Evolution of Coronal Holes, June 1973--January 1974.
Sol. Phys. 51, 377--398, 1977.

\bibitem[Bravo(1995)]{bravo1}
Bravo, S.
A Solar Scenario for the Associated Occurrence of Flares, Eruptive Prominences, Coronal Mass Ejections, Coronal Holes, and Interplanetary Shocks.
Sol. Phys. 161, 57--65, 1995.

\bibitem[Bravo(1996)]{bravo2}
Bravo, S.
A Possible Scenario for the Associated Occurrence of Flares, Prominence Eruptions, Coronal Mass Ejections, and Coronal Holes in Relation to Interplanetary Shocks.
Adv. Space Res. 17, 285--288, 1996.

\bibitem[Cole(2003)]{cole}
Cole, D.G.
Space Weather: Its Effects and Predictability.
Space Science Rev. 107, 295--302, 2003.

\bibitem[Cremades and Bothmer(2004)]{cremades1}
Cremades, H., Bothmer, V.
On the Three--Dimensional Configuration of Coronal Mass Ejections.
Astron. Astrophys. 422, 307--322, 2004.

\bibitem[Davis(1985)]{davis1}
Davis, J.M.
Small--Scale Flux Emergence and The Evolution of Equatorial Coronal Holes.
Sol. Phys. 95, 73--82, 1985.

\bibitem[de Toma(2011)]{toma}
de Toma, G.
Evolution of Cornal Holes and Implications for High--Speed Solar Wind During The Minimim Between Cycles 23 and 24.
Sol. Phys., Published online, 2011.

\bibitem[Edmondson et al.(2010)]{edmondson}
Edmondson, J.K., Antiochos, S.K., DeVore, C.R., Lynch, B.J., Zurbuchen, T.H.
Interchange Reconnection and Coronal Hole Dynamics.
Astrophys. J. 714, 517-531, 2010.

\bibitem[Echer et al.(2005)]{echer}
Echer, E., Gonzalez, W.D., Guarnieri, F.L., Lago, A. Dal, Vieira, L.E.A.
Introduction to Space Weather.
Adv. Space Res. 35, 855-865, 2005.

\bibitem[Feldman et al.(1978)]{feldman}
Feldman, W.C., Asbridge, J.R., Bame, S.J., Gosling, J.T.
Long--Term Variations of Selected Solar Wind Properties -- IMP 6, 7, and 8 Results.
Journal of Geophysical Research 83, 2177--2189, 1978.

\bibitem[Feldman et al.(1976)]{feldman2}
Feldman, W.C., Asbridge, J.R., Bame, S.J., Gosling, J.T.
High--Speed Solar Wind Flow Parameters at 1 AU
Journal of Geophysical Research 81, 5054--5060, 1976.


\bibitem[Fisk(2005)]{fisk2}
Fisk, L.A.
The Open Magnetic Flux of the Sun. I. Transport by Reconnections with Coronal Loops.
Astrophys. J. 626, 563--573, 2005.

\bibitem[Gopalswamy et al.(2003)]{gopalswamy1}
Gopalswamy, N., Shimojo, M., Lu, W., Yashiro, S., Shibasaki, K., Howard, R.A.
Prominence Eruptions and Coronal Mass Ejection: A Statistical Study Using Microwave Observations.
Astrophys. J. 586, 562--578, 2003.

\bibitem[Gopalswamy et al.(2006)]{gopalswamy2}
Gopalswamy, N., Miki\'c, Z., Maia, D., Alexander, D., Cremades, H., Kaufmann, P. Tripathi, D., Wang, Y.-M.
The Pre--CME Sun.
Space Science Rev. 123, 303--339, 2006.

\bibitem[Gopalswamy et al.(2009)]{gopalswamy3}
Gopalswamy, N., M\"akel\"a, P., Xie, H., Akiyama, S., Yashiro, S.
CME Interactions with Coronal Holes and Their Interplanetary Consequences.
Journal of Geophysical Research 114, A00A22, 2009.

\bibitem[Gonzalez et al.(1996)]{gonzalez1}
Gonzalez, W.D., Tsurutani, B.T., McIntosh, P.S., Cl\'ua de Gonzalez, A.L.
Coronal Hole--Active Region--Current Sheet (CHARCS) Association with Intense Interplanetary and Geomagnetic Activity.
Geophysical Research Letters 23, 2577-2580, 1996.

\bibitem[Harvey and Sheeley(1979)]{harvey1}
Harvey, K., Sheeley, N.R.
Coronal Holes and Solar Magnetic Fields.
Space Science Rev. 23, 139--158, 1979.

\bibitem[Harvey and Recely(2002)]{harvey2}
Harvey, K., Recely, F.
Polar Coronal Holes During Cycles 22 and 23.
Sol. Phys. 211, 31--52, 2002.

\bibitem[Hathaway(2010)]{hathaway}
Hathaway, D. H.
The Solar Cycle.
Living Rev. Solar Phys. 7, 2010.

\bibitem[Jiang et al.(2007)]{jiang1}
Jiang, Y., Liheng, Y., Li, K., Shen, Y.
Magnetic Interaction: An Erupting Filament and Remote Coronal Hole.
Astrophys. J. 667, L105--L108, 2007.

\bibitem[Kahler and Moses(1990)]{kahler1}
Kahler, S.W., Moses, D.
Discrete Changes in Solar Coronal Hole Boundaries.
Astrophys. J. 362, 728--733, 1990.

\bibitem[Kahler and Hudson(2002)]{kahler2}
Kahler, S.W., Hudson, H.S.
Boundary Structures and Changes in Long--Lived Coronal Holes.
Astrophys. J. 574, 467--476, 2002.

\bibitem[Krista et al.(2011)]{krista1}
Krista, L.D., Gallagher, P.T., Bloomfield, D.S.
Short--Term Evolution of Coronal Hole Boundaries.
Astrophys. J. Letters. 731, L26-L31, 2011.

\bibitem[Madjarska et al.(2004)]{madjarska1}
Madjarska, M.S., Doyle, J.G., van Driel--Gesztelyi, L.
Evidence of Magnetic Reconnection Along Coronal Hole Boundaries.
Astrophys. J. 603, L57--L59, 2004.

\bibitem[Madjarska and Wiegelmann(2009)]{madjarska2}
Madjarska, M.S., Wiegelmann, T.
Coronal Hole Boundaries Evolution at Small Scales: I. EIT 195$\rm{\mathring A}$ and TRACE 171$\rm{\mathring A}$ View.
Astron. Astrophys. 701, 253--259, 2009.

\bibitem[Moreno et al.(2008)]{moreno}
Moreno--Insertis, F., Galsgaard, K., Ugarte--Urra, I.
Jets in Coronal Holes: {\it HINODE} Observartions and Three-Dimensional Computer Modeling.
Astrophys. J. 673, L211--L214, 2008.

\bibitem[Munro et al.(1979)]{munro}
Munro, R.H., Gosling, J.T., Hildner, E., MacQueen, R.M., Poland, A.I., Ross, C.L.
The Association of Coronal Mass Ejection Transients with Other Forms of Solar Activity.
Sol. Phys. 61, 201--215, 1979.

\bibitem[Nolte at el.(1978)]{nolte1}
Nolte, J.T., Davis, J.M., Gerassimenko, M., Krieger, A.S., Solodya, C.V.
The Relationship Between Solar Activity and Coronal Hole Evolution.
Sol. Phys. 60, 143--153, 1978.

\bibitem[Panasenco et al.(2011)]{panasenco1}
Panasenco, O., Martin, S., Joshi, A.D., Srivastava, N.
Rolling Motion in Erupting Prominences Observed by STEREO. 
J. Amospheric and Solar--Terrestrial Physics 73, 1129--1137, 2011.

\bibitem[Pojoga and Huang(2003)]{pojoga}
Pojoga, S., Huang, T.S.
On the Sudden Disappearances of Solar Filaments and Their Relationship with Coronal Mass Ejections.
Adv. Space Res. 32, 2641--2646, 2003.

\bibitem[Priest et al.(1994)]{priest}
Priest, E.R., Parnell, C.E., Martin, S.F.
A Converging Flux Model of an X--Ray Bright Point and an Associated Canceling MAgnetic Feature.
Astrophys. J. 427, 459--474, 1974.

\bibitem[Raju et al.(2005)]{raju1}
Raju, K.P., Bromage, G.J.I., Chapman, S.A., Del Zanna, G.
Correlation Between Coronal Hole and Quiet Sun Intensities: Evidence for Continuous Reconnection.
Astron. Astrophys. 432, 341--347, 2005.

\bibitem[Raju et al.(2006)]{raju2}
Raju, K.P., Bromage, G.J.I.
EUV Line Intensitu Distribution in the Solar Atmosphere: Differences Between a Polar Coronal Hole and Its Equatorial Extension.
Astron. Astrophys. 446, 295--300, 2006.

\bibitem[Schwenn(2006)]{schwenn}
Schwenn, R.
Space Weather: The Solar Perspective.
Living Rev. Solar Phys. 3, 2006.

\bibitem[Solomon et al.(2010)]{solomon}
Solomon, S.C.; Woods, T.N.; Didkovsky, L.V.; Emmert, J.T.; Qian, L.
Anomalously Low Solar Extreme-Ultraviolet Irradiance and Thermospheric Density During Solar Minimum.
Geophys. Res. Lett. 37, L16103, 2010.


\bibitem[Subramanian(2010)]{subramanian}	
Subramanian, S., Madjarska, M.S., Doyle, J.G.
Coronal Hole Boundaries Evolution at Small Scales. II. XRT View. Can Small--Scale Outflows at CHBs Be a Source of the Slow Solar Wind.
Astron. Astrophys. 516, A50--A57, 2010.

\bibitem[Taliashvili et al.(2008)]{taliashvili1}
Taliashvili, L., Mouradian, Z., P\'aez, J.
Dynamic Disappearance of Prominences and Their Geoeffectiveness.
Geof\'\i sica Internacional 47, 279--285, 2008.

\bibitem[Taliashvili et al.(2009)]{taliashvili2}
Taliashvili, L., Mouradian, Z., P\'aez, J.
Dynamic and Thermal Disappearance of Prominences and Their Geoeffectiveness.
Sol. Phys. 258, 277--295, 2009.

\bibitem[Thompson et al.(2000)]{thompson}	
Thompson, B.J.; Cliver, E.W.; Nitta, N.; Delann\'e, C.; Delaboudini\`ere, J.-P.
Coronal Dimmings and Energetic CMEs in April-May 1998.
Geophys. Res. Lett. 27, 1431--1434, 2000.

\bibitem[Tsurutani et al.(2011)]{tsurutani}
Tsururani, B.T., Echer, E., Gonzalez, W.D.
The Solar and Interplanetary Causes of the Recent Minimum in Geomagnetic Activity (MGA23): a Combination of Midlatitude Small Coronal Holes, Low IMF $B_Z$ Variances, Low Solar Wind Speeds and Low Solar Magnetic Fields.
Ann. Geophys. 29, 839--849, 2011.



\bibitem[Waldmeier(1975)]{waldmeier1}
Waldmeier, M.
The Coronal Hole at The 7 March 1970 Solar Eclipse.
Sol. Phys. 40, 351--358, 1975.

\bibitem[Wang et al.(1996)]{wang1}
Wang, Y.-M., Hawley, S.H., Sheeley, N.R.
The Magnetic Natura of Coronal Holes.
Science 271, 464--469, 1996.

\bibitem[Wang and Sheeley(2004)]{wang2}
Wang, Y.-M., Sheeley, N.R.
Footpoint Switching and the Evolution of Coronal Holes.
Astrophys. J. 612, 1196--1205, 2004.

\bibitem[Wang(2009)]{wang3}
Wang, Y.-M.
Coronal Holes and Open Magnetic Flux.
Space Science Rev. 144, 383--399, 2009.

\bibitem[Wang and Sheeley(1993)]{wang4}
Wang, Y.-M., Sheeley, N.R.
Understanding the Rotation of Coronal Holes.
Astrophys. J. 414, 916--927, 1993.

\bibitem[Wiegelmann and Solanki(2004)]{wiegelmann}
Wiegelmann, T., Solanki, S.K.
Why Are Coronal Holes Indistinguishable From the Quiet Sun in Transition Region Radiation?
SOHO 15 Coronal Heating 575, 35--40, 2004.

	

\end{thebibliography}
\end{document}